\documentclass[12pt]{article}
\pdfoutput=1
\usepackage{graphicx,amssymb,amsmath}
\usepackage{epsfig}
\usepackage{color,graphicx}
\usepackage{cite}
\usepackage{amssymb,amsmath}
\newcommand\fverb{\setbox\pippobox=\hbox\bgroup\verb}
\newcommand\fverbdo{\egroup\medskip\noindent%
                              \fbox{\unhbox\pippobox}\ }
\newcommand\fverbit{\egroup\item[\fbox{\unhbox\pippobox}]}
\newbox\pippobox

\newcommand{\beq} {\begin{equation}}
\newcommand{\eeq} {\end{equation}}
\newcommand{\beqa} {\begin{eqnarray}}
\newcommand{\eeqa} {\end{eqnarray}}

\newcommand{\be}{\begin{equation}}
\newcommand{\ee}{\end{equation}}
\newcommand{\bea}{\begin{eqnarray}}
\newcommand{\eea}{\end{eqnarray}}

\begin{document}
 
\begin{flushright}
HIP-2014-16/TH\\
CCTP-2014-13\\
CCQCN-2014-33  
\end{flushright}

\begin{center}

\centerline{\Large {\bf Gravity dual of spin and charge density waves}}

\vspace{8mm}

\renewcommand\thefootnote{\mbox{$\fnsymbol{footnote}$}}
Niko Jokela,${}^{1,2}$\footnote{niko.jokela@helsinki.fi}
Matti J\"arvinen,${}^3$\footnote{mjarvine@physics.uoc.gr} and
Matthew Lippert${}^4$\footnote{M.S.Lippert@uva.nl}

\vspace{4mm}
${}^1${\small \sl Department of Physics} and ${}^2${\small \sl Helsinki Institute of Physics} \\
{\small \sl P.O.Box 64} \\
{\small \sl FIN-00014 University of Helsinki, Finland} 

\vspace{2mm}
${}^3${\small \sl Crete Center for Theoretical Physics} \\
{\small \sl Department of Physics} \\
{\small \sl University of Crete} \\
{\small \sl 71003 Heraklion, Greece} 

\vspace{2mm}
\vskip 0.2cm
${}^4${\small \sl Institute for Theoretical Physics} \\
{\small \sl University of Amsterdam} \\
{\small \sl 1090GL Amsterdam, Netherlands} 

\end{center}

\vspace{8mm}

\setcounter{footnote}{0}
\renewcommand\thefootnote{\mbox{\arabic{footnote}}}

\begin{abstract}
\noindent
At high enough charge density, the homogeneous state of the D3-D7' model is unstable to fluctuations at nonzero momentum.  We investigate the end point of this instability, finding a spatially modulated ground state, which is a charge and spin density wave.  We analyze the phase structure of the model as a function of chemical potential and magnetic field and find the phase transition from the homogeneous state to be first order, with a second-order critical point at zero magnetic field.
\end{abstract}

\newpage
\tableofcontents


\section{Introduction}

The ground states of many condensed matter systems feature spontaneously broken spatial symmetries.  Modulated states, such as charge and spin density waves, break translation symmetry and some or all of the symmetries of the underlying lattice.  Systems that exhibit these phases, such as doped Mott insulators, are typically characterized by competing types of order and often have complicated phase diagrams.  The pseudogap regime of cuprate superconductors, for example, features a variety of symmetry-breaking phases.

Gauge/gravity duality has proven to be a powerful tool for investigating strongly-coupled systems, such as those arising in many interesting condensed matter systems.  Although most of the focus has been on computationally simpler homogeneous and isotropic systems, in recent years symmetry-breaking instabilities and states with broken symmetry have received increasing attention.  One motivation to consider spatially inhomogeneous systems is that breaking spatial symmetries can affect interesting physical quantities.  For example, translational symmetry implies momentum conservation which automatically leads to a divergent DC conductivity.  To obtain physically sensible results, translational symmetry must be broken.

One common approach is to explicitly break the symmetries with a spatially-dependent source, such as a periodic chemical potential, as in \cite{Horowitz:2012ky, Horowitz:2012gs,Erdmenger:2013zaa, Aperis:2010cd,Hutasoit:2011rd,Ganguli:2012up,Hutasoit:2012ib,Ganguli:2013oya,Ling:2013aya,Arean:2013mta,Ling:2013nxa,Arean:2014oaa}.  Alternatively, homogeneity can be preserved and instead Lorentz invariance can be explicitly broken, as in \cite{Vegh:2013sk, Davison:2013jba, Andrade:2013gsa, Amoretti:2014zha,Gouteraux:2014hca,Taylor:2014tka,Donos:2014oha}.
This is useful for studying the properties of spatially ordered systems but does not say much about how the symmetry became broken in the first place.


In many cases, one is driven to investigate inhomogeneous states because the naive homogeneous state turns out not to be the ground state.
Investigations have found a range of examples in which homogeneous states suffer from instabilities due to large charge densities or magnetic fields \cite{Nakamura:2009tf, Ooguri:2010kt, Donos:2011bh, Donos:2011qt, Bu:2012mq, Rozali:2013ama}.  Similar instabilities have been found in models of holographic QCD, such as the Sakai-Sugimoto model, \cite{Domokos:2007kt, Ooguri:2010xs, Chuang:2010ku, Bayona:2011ab, BallonBayona:2012wx}, as well as in certain (2+1)-dimensional brane models \cite{Bergman:2011rf, Jokela:2012vn, Jokela:2012se}. 
These examples all share a common core mechanism, the instability of Maxwell-Chern-Simons theory with a constant electric field, as explained in \cite{Nakamura:2009tf}.
However, \cite{Donos:2013gda} presented a model with translational symmetry breaking but which preserved both parity and time reversal invariance.
A lot of work has gone into the difficult problem of finding the inhomogeneous end states of such instabilities \cite{Ooguri:2010kt, Ooguri:2010xs, Bayona:2011ab, Donos:2012gg, BallonBayona:2012wx, Bu:2012mq, Rozali:2012es, Donos:2012yu,Donos:2013wia, Withers:2013loa, Withers:2013kva, Rozali:2013ama, Ling:2014saa, Withers:2014sja}. 
Other interesting examples where the inhomogeneous solution is topologically nontrivial appeared in \cite{Keranen:2009ss,Keranen:2009re,Keranen:2009vi,Keranen:2010sx}.

In this paper, we will find the ground state of the D3-D7' probe-brane model. 
The homogeneous phase of this system has been analyzed in great detail, both in the metallic, black hole-embedding phase \cite{Bergman:2010gm,Bergman:2011rf,Jokela:2012vn} 
and in the quantum Hall, Minkowski-embedding phase \cite{Jokela:2010nu,Jokela:2013hta,Jokela:2014wsa}.\footnote{See also analogous results \cite{Jokela:2011sw,Jokela:2012se} in the cousin D2-D8' model \cite{Jokela:2011eb}, 
which in many respects is similar to the present case, but also has some interesting differences. The analysis carried out in the present article should find its way to the D2-D8' case, too.}
A fluctuation analysis \cite{Bergman:2011rf} showed that, above a critical charge density, the homogeneous black hole phase is unstable.  The tachyonic modes responsible for this instability have nonzero momentum, suggesting the system decays to an inhomogeneous ground state.
This instability, however, is mitigated by a nonzero mass for the fermions or an external magnetic field. 
This perturbative result motivated us to search for spatially dependent solutions and to investigate whether the expectations from the fluctuation analysis hold at the full nonlinear level.

We find explicit numerical solutions for configurations which exhibit spontaneous modulation in one
spatial dimension.  The bulk fields are entirely coupled; as a result, the corresponding inhomogeneous state is both a spin and charge density wave.\footnote{For reviews of charge and spin density waves, see \cite{Gruner1} and \cite{Gruner2}.}  The spatial frequency of this striped phase is closely related to the momenta of the tachyons of the homogeneous phase.  Comparing with the homogeneous phase, this striped phase is thermodynamically preferred at sufficiently large charge density and small magnetic field.  In general, the phase transition is first order, although there is a second-order critical point at zero magnetic field.

In the next section, we will review the construction of the D3-D7' model.  Section \ref{sec:method} will describe the pseudospectral method we employ to find our modulated solution.  In Section \ref{sec:results}, we will describe the solutions, the transition from the homogeneous phase, and the phase diagram.  We finish in Section \ref{sec:discussion} with a discussion of open questions and directions for future research.


\section{Review of the D3-D7' model}
\label{sec:review}

We begin by setting up the notation and recalling the model under study. We are interested in a holographic quantum liquid modeled 
by a probe D7-brane embedded in a black D3-brane background. The intersection of the D3-D7' system is (2+1)-dimensional, supersymmetry is completely broken, and the low-energy
excitations of the field
theory dual are purely fermionic. The model thus makes a compelling case for condensed matter applications.

The ten-dimensional background metric reads:
\be
 ds_{10}^2 = \frac{r^2}{L_{AdS}^2}\left(-h(r)dt^2+dx^2+dy^2+dz^2\right)+\frac{L_{AdS}^2}{r^2}\left(\frac{dr^2}{h(r)}+r^2d\Omega_5^2\right) \ ,
\ee
where the thermal factor is $h=1-\left(\frac{r_T}{r}\right)^4$. The metric on the internal sphere we write as an $S^2\times S^2$ fibered over an interval:
\be
 d\Omega_5^2 = d\psi^2+\cos^2\psi\left(d\theta^2+\sin^2\theta \ d\phi^2\right)+\sin^2\psi\left(d\alpha^2+\sin^2\alpha \ d\beta^2\right) \ ,
\ee
with the ranges for angles $\psi\in [0,\pi/2],$ $\theta,\alpha\in [0,\pi],$ and $\phi,\beta\in [0,2\pi]$. Recall also that the Ramond-Ramond four-form is $C^{(4)}_{txyz}=-r^4/L_{AdS}^4$, and the curvature radius
is related to the 't Hooft coupling as $L_{AdS}^2=\sqrt{4\pi g_s N}\alpha' = \sqrt{\lambda}\alpha'$. In the following we will work in units where $L_{AdS}=1$.

We will embed the D7-brane such that it spans $t,x,y$ Minkowski directions, is extended in the holographic radial direction $r$, and wraps both of the two-spheres.
We are interested in solutions which depend on $x$ and $r$, such that the embedding functions are $z=z(x,r)$ and $\psi=\psi(x,r)$. 
Moreover, to creep towards more realistic physical models, we wish to turn on all the external components of the gauge potential along the brane worldvolume: $a_{t,x,y,r}(x,r)\ne 0$.
Since the D7-brane would otherwise be unstable towards slipping off the internal space, we also turn on internal components of the gauge field, such that their field strengths are:
\bea
 2\pi\alpha' F_{\theta\phi} & = & \frac{f_1}{2}\sin\theta \\
 2\pi\alpha' F_{\alpha\beta} & = & \frac{f_2}{2}\sin\alpha \ .
\eea
The quantized constants $f_1$ and $f_2$ are proportional to the number of D5-brane fluxes diluted on the internal two-spheres.
The dynamical fields we have are $\psi,z,a_0,a_x,a_y,a_r$.  Since we are only interested in time-independent solutions, we can work in the gauge $a_r=0$.

The action, which consists of a Dirac-Born-Infeld term and a Chern-Simons term, reads:
\bea 
 S &=&  -T_7 \int d^8x\, e^{-\Phi} \sqrt{-\mbox{det}(g_{\mu\nu}+ 2\pi\alpha' F_{\mu\nu})} -\frac{(2\pi\alpha')^2T_7}{2} \int P[C_4]\wedge F \wedge F \nonumber \\
 &=& -{\cal{N}}\int dxdr\left[\sqrt G\sqrt{r^4 A+A_x+A_{xr}}+r^4 f_1 f_2 z'-2 c(\psi)(a'_0\partial_x a_y-\partial_x a_0 a'_y)\right] \ ,
\eea
where the prime denotes differentiation with respect to $r$, 
\bea
 G & = & \left(f_1^2+4\cos^4\psi\right)\left(f_2^2+4\sin^4\psi\right) \\ 
 A & = & 1+hr^4z'^2+hr^2\psi'^2-a'^2_0+ha'^2_x+ha'^2_y \\
 A_x & = & -\frac{1}{h}\partial_x a_0^2+\partial_x a_y^2+r^4\partial_x z^2+r^2\partial_x\psi^2 \\
 A_{xr} & = & -\partial_x a^2_0 a'^2_y-r^4z'^2\partial_x a^2_0-r^2\psi'^2\partial_x a^2_0-a'^2_0\left(r^4\partial_x z^2+r^2\partial_x\psi^2+\partial_x a_y^2\right) \nonumber \\
 & & +hr^2\Bigg[r^2z'^2\partial_x a_y^2+\psi'^2\partial_x a_y^2+r^4\psi'^2\partial_x z^2+r^2 a'^2_y\partial_x z^2+r^4z'^2\partial_x\psi^2+ \nonumber \\
 & & a'^2_y\partial_x\psi^2 -2r^4z'\psi'\partial_x z\partial_x\psi-2a'_y\partial_x a_y(r^2z'\partial_x z+\psi'\partial_x\psi) \Bigg] \nonumber \\
 & & +2a'_0 \partial_x a_0\left(a'_y\partial_x a_y+r^4z'\partial_x z+r^2\psi'\partial_x\psi \right) \ ,
\eea
and
\be
 c(\psi) = \psi-\frac{1}{4}\sin(4\psi)-\psi_\infty+\frac{1}{4}\sin(4\psi_\infty) \ .
\ee
The overall constant factor of the action is ${\cal{N}} =4\pi^2T_8V_{1,1}$, where $V_{1,1}$ is the spacetime volume in the homogeneous $t$ and $y$ directions. 
Notice that $a_x$ only appears once, as a single term $ha_x'^2$ in $A$, which means that there is a trivial constant (both in $x$ and $r$) solution for it and it decouples from the rest of the dynamics; we can thus drop $a_x$ in what follows.
Before proceeding to solve the equations of motion that follow from the above action, let us make a coordinate transformation and in particular absorb one scale.

The model has several parameters. These correspond to some boundary values ($r\to\infty$) of the bulk fields $\psi,z,a_0,a_y$, as well as the fluxes $f_1$ and $f_2$,
together with the parameter $r_T$ of the background which is proportional to the temperature. Not all the parameters are independent, however.  It was shown in \cite{Bergman:2010gm} that $f_1$ and $f_2$ fix $\psi(r\to\infty)\equiv \psi_\infty$ via the relation
\be
\label{eq:f1f2psiinfty}
 f_1^2\sin^2\psi_\infty-f_2^2\cos^2\psi_\infty+4\sin^2\psi_\infty\cos^2\psi_\infty\left(\cos^2\psi_\infty-\sin^2\psi_\infty\right)=0 \ .
\ee
The fluxes also determine the subleading exponent of the field $\psi\sim\psi_\infty + A r^{\Delta_\pm}$:
\be
 \Delta_\pm = -\frac{3}{2}\pm \frac{1}{2}\sqrt{9+16\frac{f_1^2+16\cos^6\psi_\infty-12\cos^4\psi_\infty}{f_1^2+4\cos^6\psi_\infty}} \ ,
\ee
where we have used (\ref{eq:f1f2psiinfty}) to eliminate $f_2$. We note that, in order for the model to be perturbatively stable, one needs large enough flux to have real exponents $\Delta_\pm$.
In this paper we will only consider the case $f_1=f_2=1/\sqrt{2}$ which implies that $\psi_\infty=\pi/4$ and also that the anomalous mass dimension of the fermions is vanishing:
\be
\psi\sim \frac{\pi}{4}+\frac{m_\psi}{r}-\frac{c_\psi}{r^2} \ ,
\ee
where $m_\psi$ is proportional to the quark mass and $c_\psi$ is proportional to the condensate.  
The field $\psi$ plays the role of an axion in this model and couples linearly to the magnetic field, as will become explicit in the Chern-Simons term of the action (\ref{eq:uaction}) below. 
As was shown in \cite{Jokela:2012vn}, the magnetization is directly related to the radial profile of $\psi$.
One of the main results of this paper is that the fermion bi-linear $c_\psi$ will be modulated, resulting in a modulated magnetization, which we identify as a spin density wave.\footnote{In this probe brane system, there is no explicit holographic representation of the fermion spins.  However, we can use the magnetization density as a proxy for the spin density.}
For simplicity, we will consider massless fermions in this paper, $m_\psi = 0$.

We can freely set the boundary values of $z$ and $a_y$ to zero since neither $z$ nor $a_y$ enter in the equations of motion without derivatives and they do not also have IR constraints. 
This is unlike for the case of $a_0$, whose IR value
has to vanish at the horizon, leaving the UV boundary value $a_0(r\to\infty)\equiv \mu$ as a physical parameter of the theory. We will specify the UV boundary conditions explicitly for the different fields
in App. \ref{app:boundaryconditions}.

We have a freedom to scale out one parameter of the model and for this we choose to pick the temperature. We will introduce a new compact radial variable
\be
 u = \frac{r_T}{r} \ . 
\ee
To get rid off explicit dependence of $r_T$ factors everywhere, we  introduce notation with hats as follows:
\be
 \hat x^\mu = \frac{x^\mu}{r_T}\quad ,\quad \hat z=\frac{z}{r_T}\quad ,\quad \hat a_\mu = \frac{a_\mu}{r_T} \ .
\ee
Essentially, all the parameters are read in the units of the horizon radius. 

The action in the new radial coordinate reads
\bea 
 S & = & -{\cal{N}}_T\int d\hat x du \ u^{-2}\Bigg[\sqrt G\sqrt{u^{-4} \hat A+\hat A_{x}+\hat A_{xu}}-u^{-2}f_1 f_2 \hat z' \\
  & & +2 c(\psi)u^2( \hat a'_0\partial_{\hat x}\hat a_{y}-\partial_{\hat x}\hat a_{0}\hat a'_{y})\Bigg] \ ,\label{eq:uaction}
\eea
where\footnote{Note that we have dropped the decoupled field $\hat a_{x}$.}
\bea
 G & = & \left(f_1^2+4\cos^4\psi\right)\left(f_2^2+4\sin^4\psi\right) \\ 
 \hat A & = & 1+h u^2 \psi'^2+h \hat z'^2-u^4 \hat a'^2_{0}+h u^4\hat a'^2_{y} \\
\label{Axexp} 
\hat A_{x} & = & -\frac{1}{h}\partial_{\hat x}\hat a_{0}^2+\partial_{\hat x}\hat a_{y}^2+u^{-4}\partial_{\hat x}\hat z^2+u^{-2}\partial_{\hat x}\psi^2 \\
 \hat A_{xu} & = & -u^4\partial_{\hat x}\hat a^2_{0}\hat a'^2_{y}-\hat z'^2\partial_{\hat x}\hat a^2_{0}-u^2\psi'^2\partial_{\hat x}\hat a^2_{0}-u^{4}\hat a'^2_{0}\left(u^{-4}\partial_{\hat x}\hat z^2+u^{-2}\partial_{\hat x}\psi^2+\partial_{\hat x} \hat a_{y}^2\right) \nonumber \\
 & & +h \hat z'^2\partial_{\hat x}\hat a_{y}^2+h u^2\psi'^2\partial_{\hat x}\hat a_{y}^2+h u^{-2}\psi'^2\partial_{\hat x}\hat z^2+h\hat a'^2_{y}\partial_{\hat x} z^2+h u^{-2}\hat z'^2\partial_{\hat x}\psi^2+ \nonumber \\
 & & h u^2 \hat a'^2_{y}\partial_{\hat x}\psi^2 -2h u^{-2}\hat z'\psi'\partial_{\hat x}\hat z\partial_{\hat x}\psi-2 h\hat a'_{y}\partial_{\hat x} \hat a_{y}(\hat z'\partial_{\hat x}\hat z+u^2\psi'\partial_{\hat x}\psi) \nonumber \\
 & & +2u^2\hat a'_{0} \partial_{\hat x}\hat a_{0}\left(u^2\hat a'_{y}\partial_{\hat x}\hat a_{y}+u^{-2}\hat z'\partial_{\hat x}\hat z+\psi'\partial_{\hat x}\psi \right) \ ,
\eea
and
\bea
 c(\psi) & = & \psi-\frac{1}{4}\sin(4\psi)-\psi_\infty+\frac{1}{4}\sin(4\psi_\infty) \\
 h & = & 1-u^4 \ .
\eea
The new overall factor is ${\cal{N}}_T=r_T^2{\cal{N}}$. The prime now denotes differentiation with respect to $u$.

The equations of motion that we will study in this article are the Euler-Lagrange equations for the fields $\psi=\psi(\hat x,u),\ \hat z=\hat z(\hat x,u),\ \hat a_{0}=\hat a_{0}(\hat x,u)$, and $\hat a_{y}=\hat a_{y}(\hat x,u)$, which follow from (\ref{eq:uaction}). To streamline
the discussion and because they are quite lengthy, we will not write the equations of motion down explicitly.


\section{Solution by pseudospectral method}
\label{sec:method}

We will now briefly sketch the techniques we used to solve the equations of motion, which are a coupled, nonlinear, second-order PDE system.  A more detailed description can be found in Appendix \ref{app:method}.

We employed a pseudospectral method (see, {\emph{e.g.}}, \cite{boyd}), in which the difficult partial differential equations are rendered into a more tractable system of algebraic equations. Most of the runs could be performed on a personal computer, but we ended up using a supercomputing cluster for longer runs in order to achieve sufficient accuracy to analyze the phase structure. 

The first step is to determine the correct set of boundary conditions.  The boundary conditions imposed must be consistent with the equations of motion and must fix all constants of integration. We are interested in the {\emph{spontaneous}} breaking of translation invariance rather than breaking it explicitly. 
Therefore, the nonnormalizable terms in the UV remain as unmodulated free parameters.  The normalizable terms will be, in general, modulated, 
and their values are fixed by requiring regularity in the IR.

In the UV, at $u=0$, we fix the unmodulated sources as follows (see Appendix \ref{app:boundaryconditions} for more detail): 
\bea
 \partial_u \psi(\hat x,0) &=& \hat m_\psi \\
 \hat a_{0}(\hat x,0) &=& \hat\mu\\
 \hat a_{y}(\hat x,0) &=& \hat b\hat x \ .
\eea
where $\hat\mu$ is the (reduced) chemical potential for the fermions of mass $\hat m_\psi,$ and $\hat b$ is the (reduced) external magnetic field:\footnote{Note that, although $\hat a_y$ has a linear dependence on $\hat x$, the resulting magnetic field $\hat f_{xy} = \hat b$ is homogeneous and isotropic.}
\bea
 \hat m_\psi & = & \frac{m_\psi}{r_T} \\ 
 \hat\mu & = & \frac{\mu}{r_T} \\
 \hat b & = & \frac{b}{r^2_T} \ .
\eea

At the horizon $u \to 1$, we demand that the solutions are regular and finite. One particular requirement for the regularity is that $\hat a_0$ is constant in the IR but in fact we need to require this constant to vanish:
\be
 \hat a_{0}(\hat x,1) = 0 \ ,
\ee
which ensures that $\hat a$ is also a well-defined one-form as the thermal circle pinches off.

Finally, we need to specify the boundary conditions in the $x$-direction.  We are looking for spatially modulated solutions, so it is natural to impose periodic boundary conditions; that is, $\psi(\hat x) = \psi(\hat x+\hat L)$, etc, where $L = r_T \hat L$ is the spatial period. However, it is not obvious a priori what is the correct period $\hat L$ (or equivalently frequency $\hat k =\frac{2 \pi}{\hat L}$) to choose.  Our approach is to find solutions for a range of spatial periods, and by comparing their free energies, determine which wavelength is preferred by the system.  

Having fixed the boundary conditions, we proceed to turn the system of differential equations of motion into a system of algebraic equations. To do this, we implement the standard pseudospectral method based on a Fourier series having $N_x$ terms in the $\hat x$-direction and an expansion in the Chebyshev polynomials having $N_u$ terms in the $u$-direction. The values of $\psi$, $\hat z$, $\hat a_{0}$, and $\hat a_{y}$ at the $\mathcal O(N_x N_u)$ collocation points are chosen as the variables, and the derivatives are computed in the pseudospectral approximation, i.e., from the Fourier and Chebyshev series with $\mathcal O(N_x N_u)$ terms which exactly match with the variables at the collocation points. We then insert these numbers into the equations of motion.  By demanding that the equations are satisfied at the collocation points, we obtain $\mathcal O(N_x N_u)$ algebraic equations for the values of the functions which we then solve numerically. See Appendix~\ref{app:method} for more details. The accuracy 
converges exponentially 
with increasing $N_x$ and $N_u$. We tested the code up to $N_x = 38$ and $N_u = 40$, which yielded an accuracy of at least $10^{-9}$ for all functions.


\section{Results}
\label{sec:results}

\subsection{Striped solutions without a magnetic field}

Having set up the numerical method for solving a system of coupled PDEs, we can vary the parameters $\hat \mu$, $\hat L$, and $\hat b$ and look for striped solutions.  We consider massless fermions, so $\hat m_\psi = 0$.  For simplicity, we begin with the restricted case of zero magnetic field and generalize to $\hat b \not= 0$ in Sec.~\ref{sec:magneticstripes}.  

\begin{figure}[!ht]
\center
\includegraphics[width=0.45\textwidth]{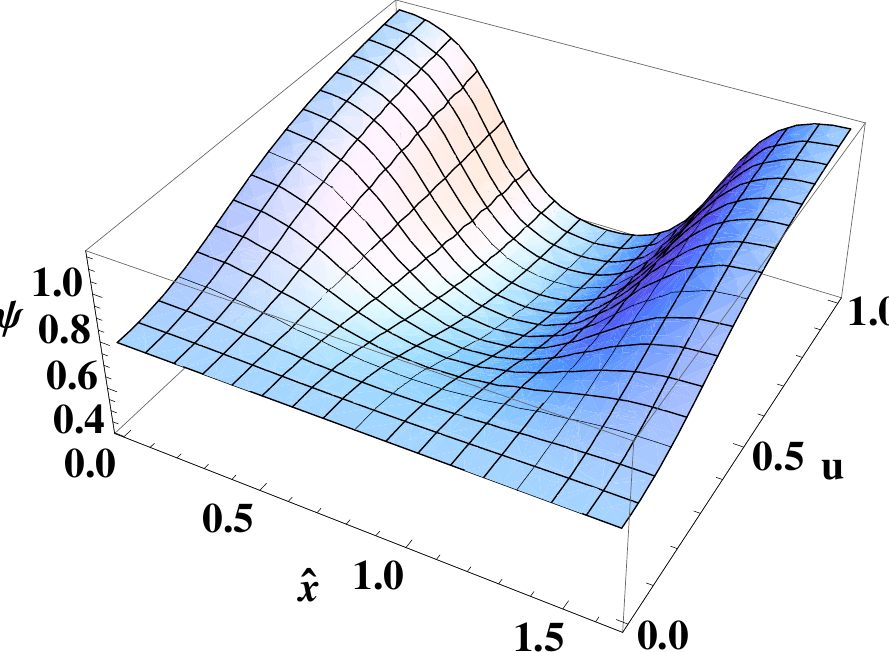} \ \ 
\includegraphics[width=0.45\textwidth]{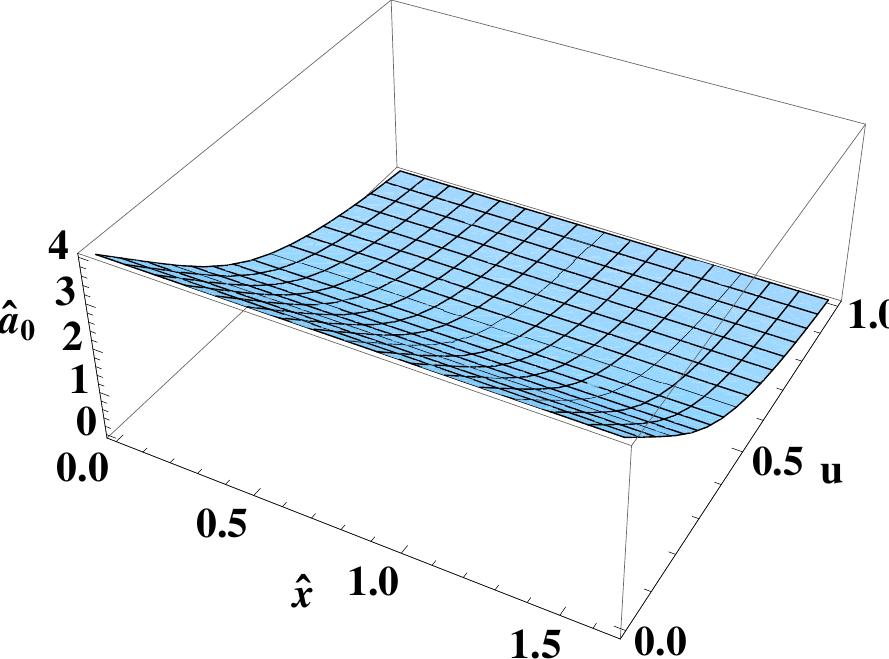}\\
\includegraphics[width=0.45\textwidth]{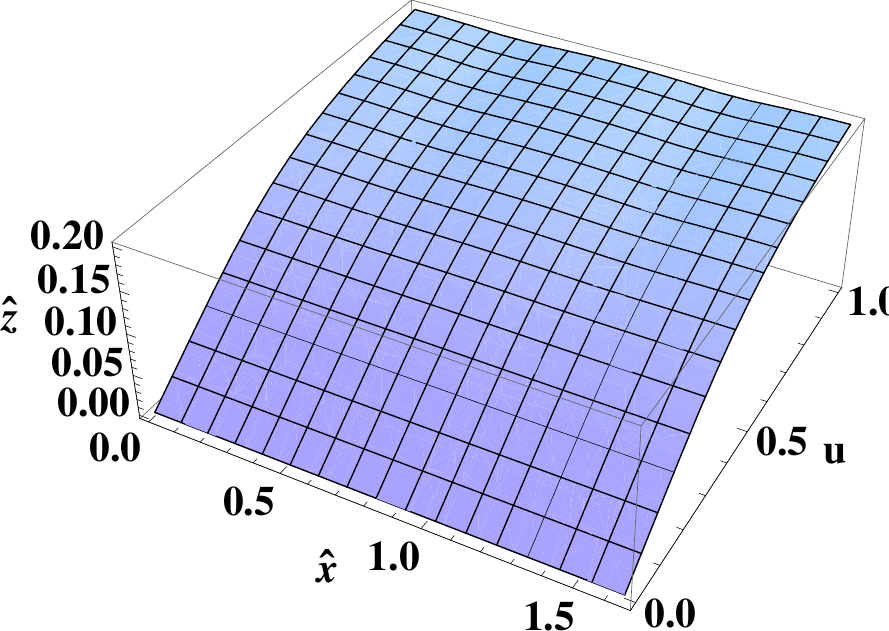} \ \
\includegraphics[width=0.45\textwidth]{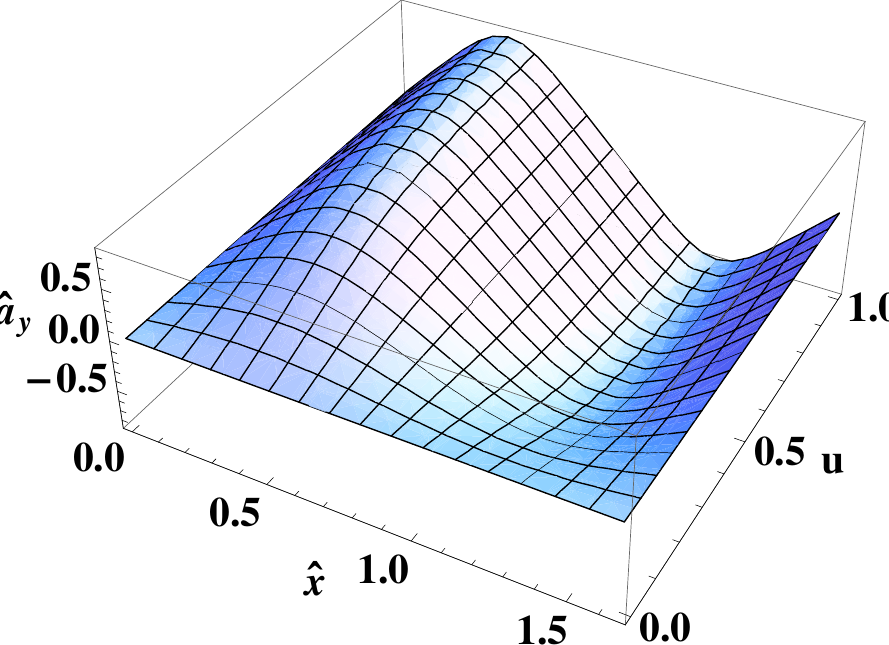}
\caption{A representative modulated solution: $\psi$ (top left), $\hat a_0$ (top right), $\hat z$ (bottom left), and $\hat a_y$ (bottom right) at $\hat L = 1.646$, and $\hat\mu = 4$.  This is the minimum energy solution, as shown in Fig.~\ref{fig:Omega_at_zero_b}, with $\hat k= \hat k_0= 3.815$.}
\label{fig:solution1}
\end{figure}

\begin{figure}[!ht]
\center
\includegraphics[width=0.45\textwidth]{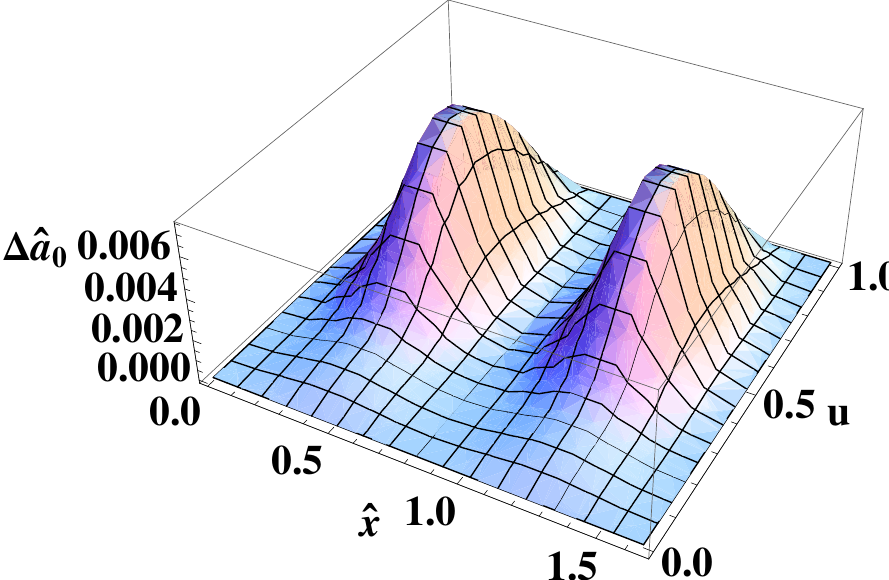} \ \ 
\includegraphics[width=0.45\textwidth]{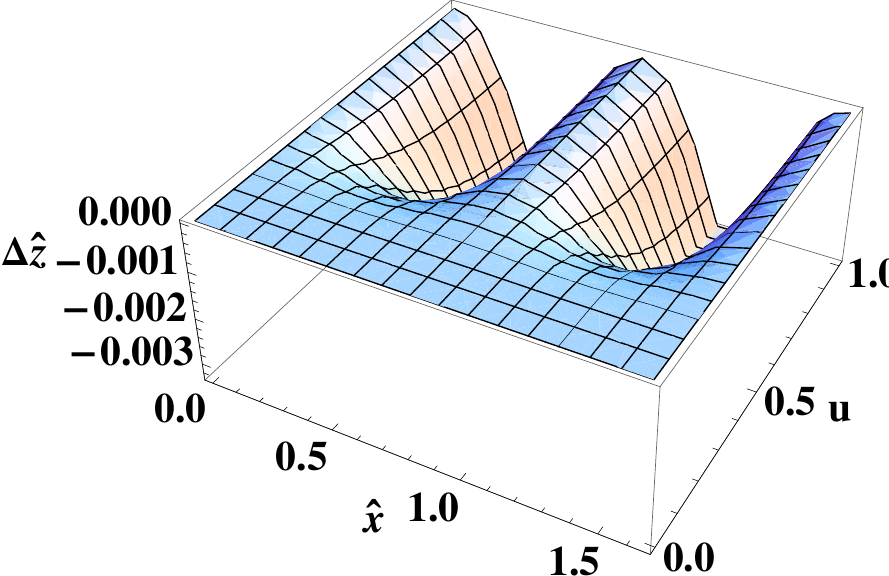}
\caption{The modulations (left) $\Delta \hat a_0(\hat x,u) = \hat a_0(\hat x,u) - \hat a_0(0,u)$ and (right) $\Delta \hat z(\hat x,u) = \hat z(\hat x,u) - \hat z(0,u)$  at $\hat L = 1.646$ and $\hat\mu = 4$.}
\label{fig:solution2}
\end{figure}

A representative striped solution is shown in Fig.~\ref{fig:solution1}. All the fields are modulated: $\psi$ and $\hat a_y$ have periodicity equal to the periodicity imposed on the space $\hat L_{soln} = \hat L$ but are $\pi/2$ out of phase with each other; $\hat a_0$ and $\hat z$ are modulated with frequency $2\hat k$.  Because of the charge density and wrapped internal flux, the bulk fields $\hat a_0$ and $\hat z$ have nontrivial homogeneous profiles; in Fig.~\ref{fig:solution2}, we have subtracted out this homogeneous background to highlight the modulation.  By construction, the modulation goes to zero near the boundary $u \to 0$, indicating that the translation symmetry is being spontaneously broken.

The action \eqref{eq:uaction} is covariant under the parity transformation $\hat x \to -\hat x$, and this symmetry is preserved in all the solutions we have found. When the magnetic field is zero there is also another reflection symmetry. 
Similarly the solutions with $\hat b=0$ retain this second symmetry under reflections $\hat x \to \frac{\hat L}{2} - \hat x$. (See App. \ref{app:boundaryconditions} for more details.)

To determine the preferred modulation frequency, we compare the free energies of solutions of varying $\hat L$, with $\hat\mu$ held fixed.  As we are working at fixed $\hat\mu$, that is, in the grand canonical ensemble, the relevant free energy is the grand canonical potential, which is defined by the Euclidean action \eqref{eq:uaction} evaluated on a solution to the equations of motion:
\be\label{eq:onshellaction}
\Omega(\hat\mu,\hat b) = \frac{1}{\mathcal N_T}  \left. S^{E}[\psi(u), \hat a_\mu(u), \hat z(u)] \right|_{on-shell} \ .
\ee
This divergent expression (\ref{eq:onshellaction}) is regulated using the standard holographic renormalization; the required counterterms are explicitly written in \cite{Bergman:2010gm}.  
Note that the range of the $\hat x$ integral in the action is $\hat L$.  In order to compare solutions with different periodicities $\hat L$, the quantities we need to compare are the average grand potential densities $\Omega/\hat L$.  Bear in mind that, while we are artificially fixing the spatial periodicity $\hat L$ by hand here, in the actual system the periodicity is determined dynamically as the system minimizes its energy.  

\begin{figure}[!ht]
\center
\includegraphics[width=0.7\textwidth]{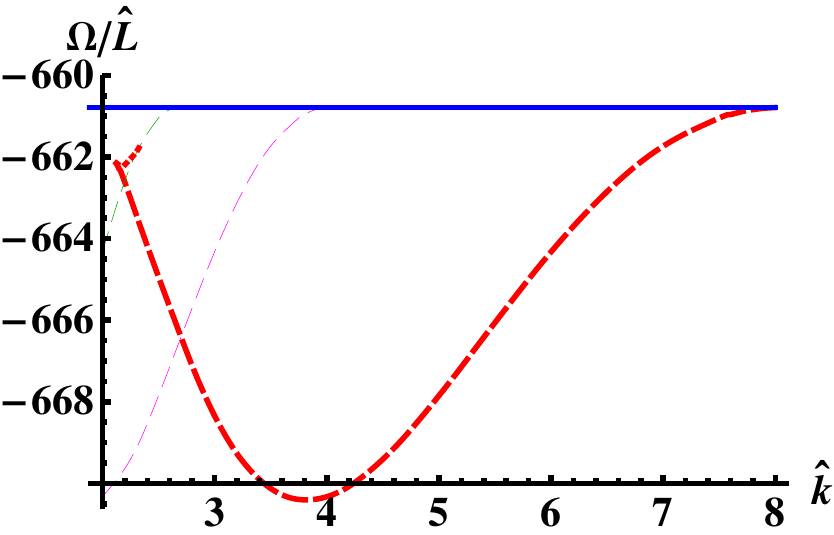}
\caption{$\Omega/\hat L$ for the various solutions at $\hat\mu = 4$. The different dashed curves correspond to branches of solutions with different numbers of oscillations $n$ within the given spatial range $\hat L$.  The red branch is $n=1$, the magenta branch is $n=2$, and green branch is $n=3$.  The short dotted red curve is a family of solutions which interpolates between the $n=1$ and $n=3$ branches.  The horizontal blue line shows $\Omega/\hat L$ for the homogeneous solution.  The frequency at which $\Omega/\hat L$ is minimized is $\hat k_0 = 3.815$.  The dashed red $n=1$ branch merges with the homogeneous solution at $\hat k_{max} = 8.05$ and with the red interpolating branch at $\hat k_{min} =2.124$.  The metastable solution on the dotted red branch has $\hat k = 2.172$.}
\label{fig:Omega_at_zero_b}
\end{figure}

In Fig.~\ref{fig:Omega_at_zero_b}, we plot $\Omega/\hat L$ as a function of $\hat k = \frac{2\pi}{\hat L}$ for a fixed $\hat\mu$.  The various curves on the plot correspond to different ratios between the periodicity of $\hat x$, which is $\hat L$, and the periodicity of the solution, denoted by $\hat L_{sol}$.  For the solution shown in Fig.~\ref{fig:solution1}, the two periods are equal, but in general there may be any integer number of modulations of the solution within the spatial period, {\it i.e.}, $\hat L = n \hat L_{sol}$ for any positive integer $n$.  In particular, if a field configuration with modulation period $\hat L_{sol}$ is a solution when $\hat L = \hat L_{sol}$, it will necessarily also be a solution for $\hat L = n \hat L_{sol}$. 

\begin{figure}[!ht]
\center
\includegraphics[width=0.7\textwidth]{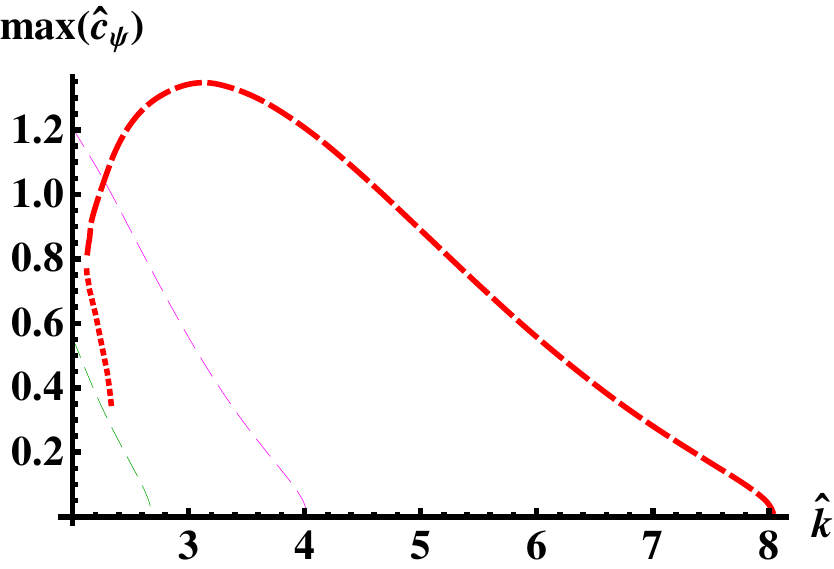}
\caption{The amplitude of modulation, illustrated by the maximum (over $\hat x$) of $\hat c_\psi$.  For the homogeneous solutions $\hat c_\psi = 0$.  The curves correspond to the branches of solutions in Fig. \ref{fig:Omega_at_zero_b}. }
\label{fig:maxcpsi_vs_k}
\end{figure}

As can be seen from Fig.~\ref{fig:Omega_at_zero_b}, there is a minimum energy solution at nonzero $\hat k= \hat k_{0}$; this is the frequency preferred by the system for a given $\hat \mu$ and is the frequency the system will take when $\hat L$ is not fixed by hand.  Forcing the system to have a higher frequency $\hat k > \hat k_0$ costs energy, and the amplitude of the modulation decreases, as shown in Fig. \ref{fig:maxcpsi_vs_k}. For sufficiently large $\hat k = \hat k_{max}$, the modulation vanishes, and the $n=1$ branch of modulated solutions merges smoothly with the homogeneous solution.  When $\hat k > \hat k_{max}$, there are no striped solutions.  

In the opposite limit, if we make the spatial periodicity large $\hat L \gg \frac{2\pi}{\hat k_0}$, the system will try to have as many oscillations as necessary so $\hat k$ is as close to $\hat k_0$ as possible.  As $\hat k$ is decreased below $\hat k_0$, the energy similarly grows and the amplitude decreases, eventually reaching a minimum $\hat k = \hat k_{min}$.  

Rather than merging with the homogeneous solution, the $n=1$ branch of solutions merges at $\hat k_{min}$ with a branch of solutions which interpolates between the $n=1$ and $n=3$ solutions (shown in red dots in Fig.~\ref{fig:Omega_at_zero_b}).  Interestingly, along this interpolating branch of solutions is a local minimum of $\Omega/\hat L$, indicating the presence of a metastable solution.

\begin{figure}[!ht]
\center
\includegraphics[width=0.7\textwidth]{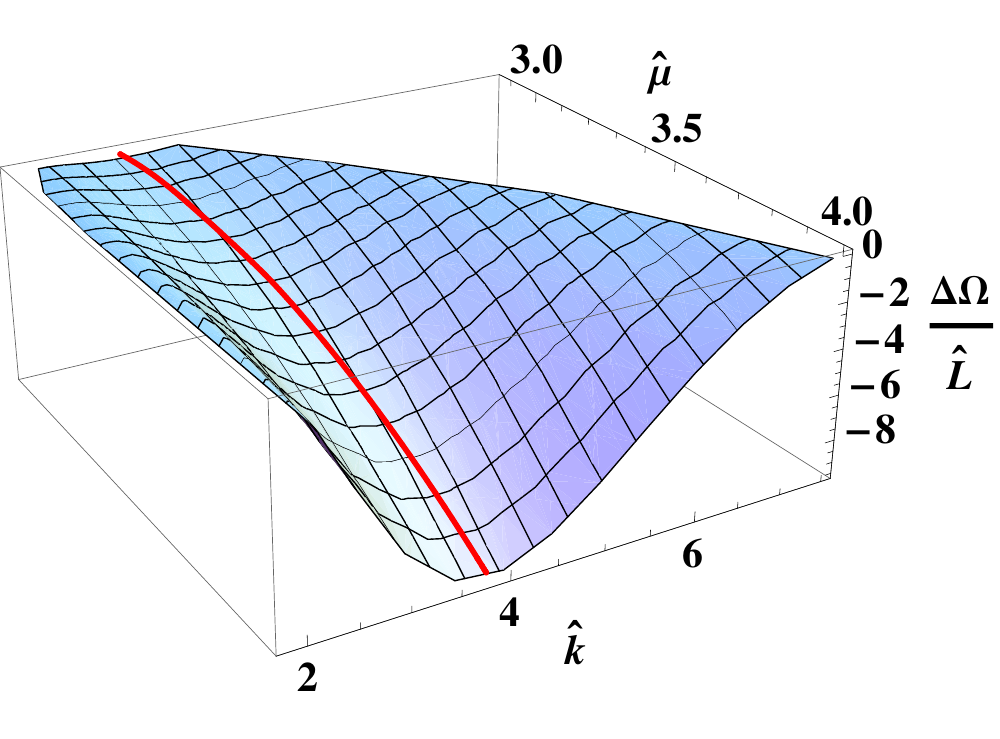}
\caption{$\Delta \Omega/\hat L$ as a function of $\hat\mu$ and $\hat k$. The red curve shows $\hat k_0(\hat\mu)$, which minimizes $\Delta \Omega/\hat L$ for each $\hat\mu$.  The critical chemical potential is $\hat\mu_c = 2.66$, and the critical frequency is $\hat k_c = 2.82$.}
\label{fig:DeltaOmega_vs_mu_at_zero_b}
\end{figure}

We now allow $\hat \mu$ to vary, and compute $\Omega/\hat L$ as a function of both $\hat\mu$ and $\hat k$.  In Fig.~\ref{fig:DeltaOmega_vs_mu_at_zero_b}, we plot the difference between the modulated and homogeneous free energies $\Delta\Omega/\hat L = (\Omega_{mod} - \Omega_{hom})/\hat L$.  For $\hat\mu$ above the critical chemical potential $\hat\mu_c$, there is a range of $\hat k$ where $\Delta\Omega/\hat L < 0$, indicating that the modulated solution is preferred.  The preferred frequency $\hat k_0(\hat\mu)$ is given by the $\hat k$ that minimizes $\Delta\Omega(\hat\mu)/\hat L$ and sets the frequency of the striped phase.  As shown in Fig.~\ref{fig:DeltaOmega_vs_mu_at_zero_b}, the striped phase appears at $\hat\mu_c$ with a nonzero critical frequency $\hat k_c = \hat k_0(\hat\mu_c)$, and $\hat k_0(\hat\mu)$ increases with $\hat\mu$.

\begin{figure}[!ht]
\center
\includegraphics[width=0.45\textwidth]{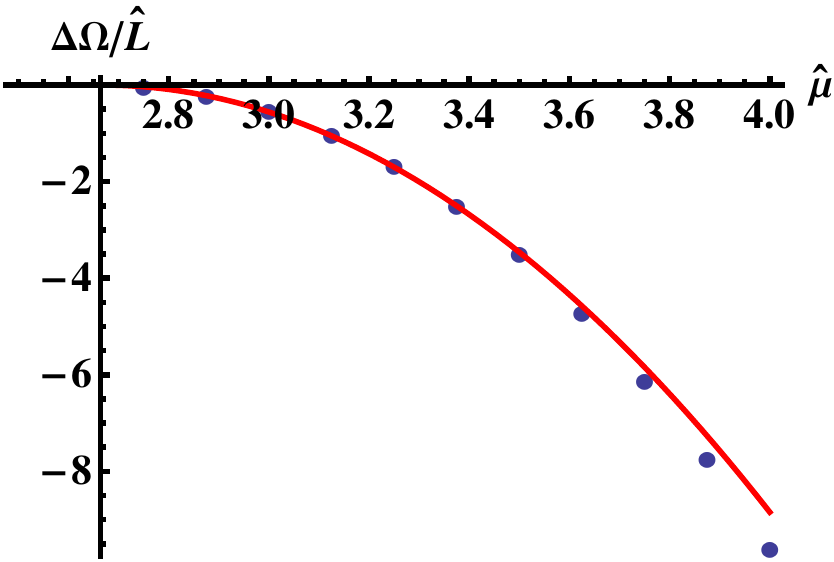}
\includegraphics[width=0.45\textwidth]{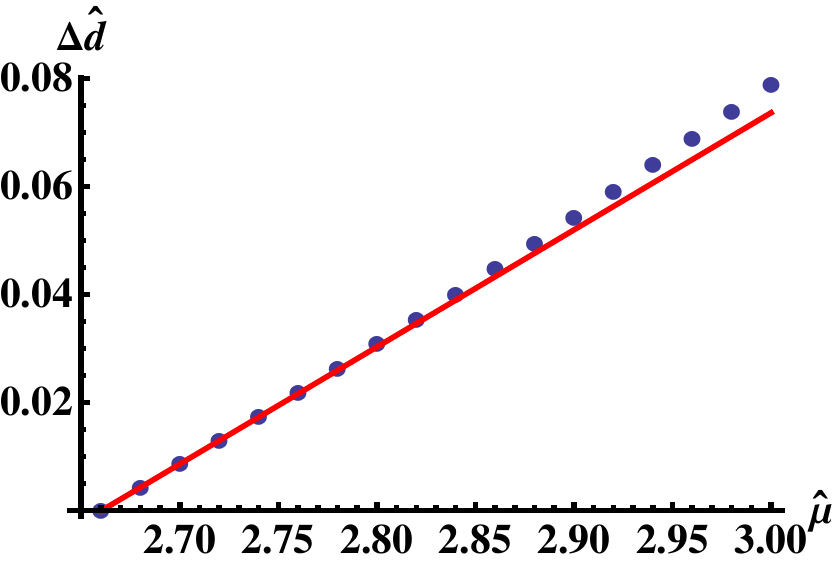}
\caption{The phase transition between the homogeneous and striped phases illustrated by the difference in free energy $\Delta \Omega(\hat k_0)/\hat L$ (left) and charge densities $\hat d$ of the two phases (right) as functions of $\hat\mu$. Left: The blue points are the numerically computed values of $\Delta \Omega(\hat k_0)/\hat L$, and the red curve is the quadratic fit $\Delta\Omega = -4.96 (\hat\mu - \hat\mu_c)^2$.  Right: The difference in density $\Delta\hat d$ between the two phases as a function of $\hat\mu$ just above the $\hat\mu_c$.  The numerical values are the blue points, and the red line is a linear fit $\Delta \hat d = 0.217 (\hat\mu - \hat\mu_c)$.}
\label{fig:Omega_and_d_at_zero_b}
\end{figure}

The transition between the homogeneous and the striped phases occurs at $\hat\mu_c$.  As $\hat\mu$ decreases to $\hat\mu_c$, $\Delta\Omega(\hat k_0, \hat\mu)/\hat L \to 0$ smoothly, as illustrated in Fig.~\ref{fig:Omega_and_d_at_zero_b}(left).  At $\hat\mu_c$, the charge density $\hat d = - \frac{\partial \Omega}{\partial \hat\mu}$ of the two phases are equal, as shown in Fig.~\ref{fig:Omega_and_d_at_zero_b}(right), indicating that the phase transition is continuous.  We can fit to very good accuracy the difference in densities of the two phases $\Delta\hat d = \hat d_{mod} - \hat d_{hom}$ near $\hat\mu_c$ to a linear function of $\hat\mu$.  We conclude from this that the phase transition is second-order.

\begin{figure}[!ht]
\center
\includegraphics[width=0.45\textwidth]{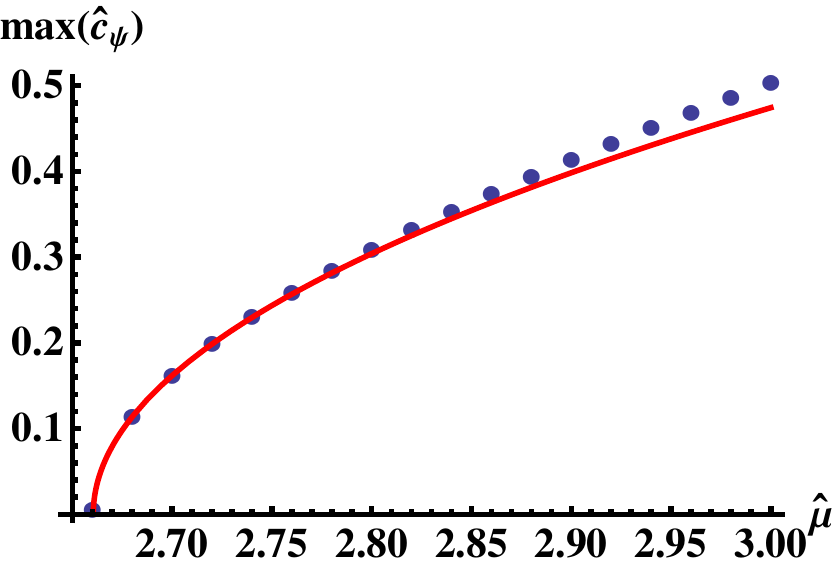}
\includegraphics[width=0.45\textwidth]{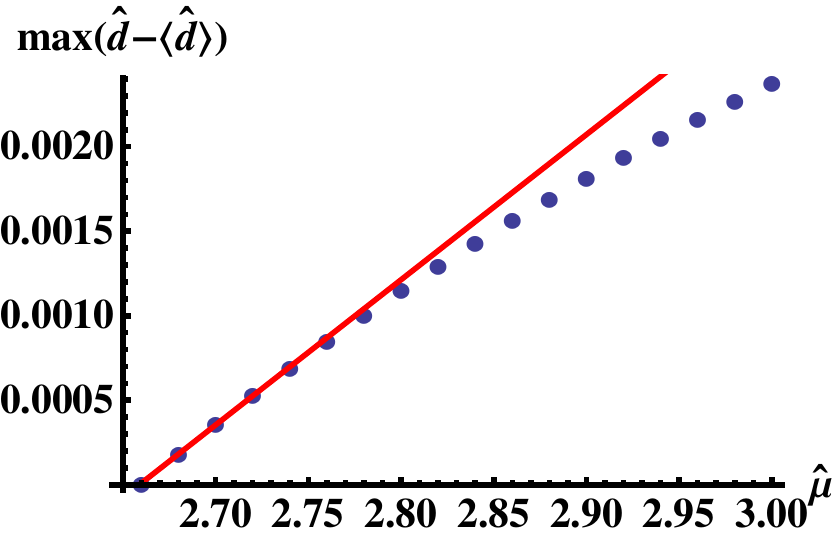}
\caption{The maximum over $\hat x$ of $\hat c_\psi$ (left) and the maximal deviation from the mean value of $\hat d$ (right), illustrating the amplitude of modulation as a function of chemical potential $\hat\mu$.  Note that the homogeneous solution has $\hat c_\psi = 0$.  The blue dots are numerically computed, and the red is the fit to ${\rm max} \ \hat c_\psi = 0.922 \sqrt{\hat\mu - \hat\mu_c}$ in the left hand plot and ${\rm max} \ (\hat d-\langle \hat d \rangle ) = 0.00859 (\hat\mu - \hat\mu_c)$ in the right hand plot.}
\label{fig:maxcpsi_vs_mu_fit_k}
\end{figure}

We can further investigate the nature of the phase transition by analyzing how the modulation disappears as the critical point is approached.  Fig. \ref{fig:maxcpsi_vs_mu_fit_k} shows the amplitude of modulation, which is well fit near the critical point by $\sqrt{\hat\mu - \hat\mu_c}$ for the amplitude of the spin density wave (or rather its proxy $\hat c_\psi$) and by a linear function for the amplitude of the charge density wave ($\hat d-\langle \hat d \rangle$).

This phase transition exactly matches the expectations obtained from the fluctuation analysis of the homogeneous phase.  As shown in \cite{Bergman:2011rf}, at a particular chemical potential $\hat\mu_{inst}$, one quasinormal mode develops a positive imaginary dispersion at nonzero momentum $\hat k_{inst}$; see Sec.~4.4 of \cite{Bergman:2011rf},\footnote{The hatted variables in \cite{Bergman:2011rf} are defined as they are here.  However, in \cite{Bergman:2011rf}, the spectrum is computed as a function of charge density $\hat d$, making it necessary to translate from $\hat d$ to $\hat\mu$ in order to make a direct comparison.} particularly Fig.~4.
The chemical potential and momentum at which the instability occurs match the critical chemical potential and spatial frequency computed here from the explicit modulated solutions: $\hat\mu_{inst} = \hat\mu_c$ and $\hat k_{inst} = \hat k_c$.

There is another lesson to be drawn. Comparing the magnitudes of the modulations of different fields, for example as depicted in Fig.~\ref{fig:maxcpsi_vs_mu_fit_k}, we see that the amplitude of the spin wave dominates over that of the charge density wave. This conforms nicely with the fluctuation analysis \cite{Bergman:2011rf}. When the fermion mass is set to zero $\hat m_\psi = 0$ at zero magnetic field strength $\hat b=0$, the fluctuation equations decouple. The tachyonic mode responsible for the modulated instability resides entirely in the sector which only mixes the embedding scalar $\delta\psi$ fluctuation and the transverse gauge field fluctuation $\delta\hat a_y$, thus suggesting only a spin density wave. Comparison of the plots in Fig.~\ref{fig:maxcpsi_vs_mu_fit_k} shows, however, that coupling to the charge density wave is present at second order in the fluctuation analysis. It is also interesting to note that the nonlinearities do not essentially ramp up the charge density wave, even further away 
from the critical point.

\subsection{Stripes in a magnetic field}
\label{sec:magneticstripes}

\begin{figure}[!ht]
\center
\includegraphics[width=0.7\textwidth]{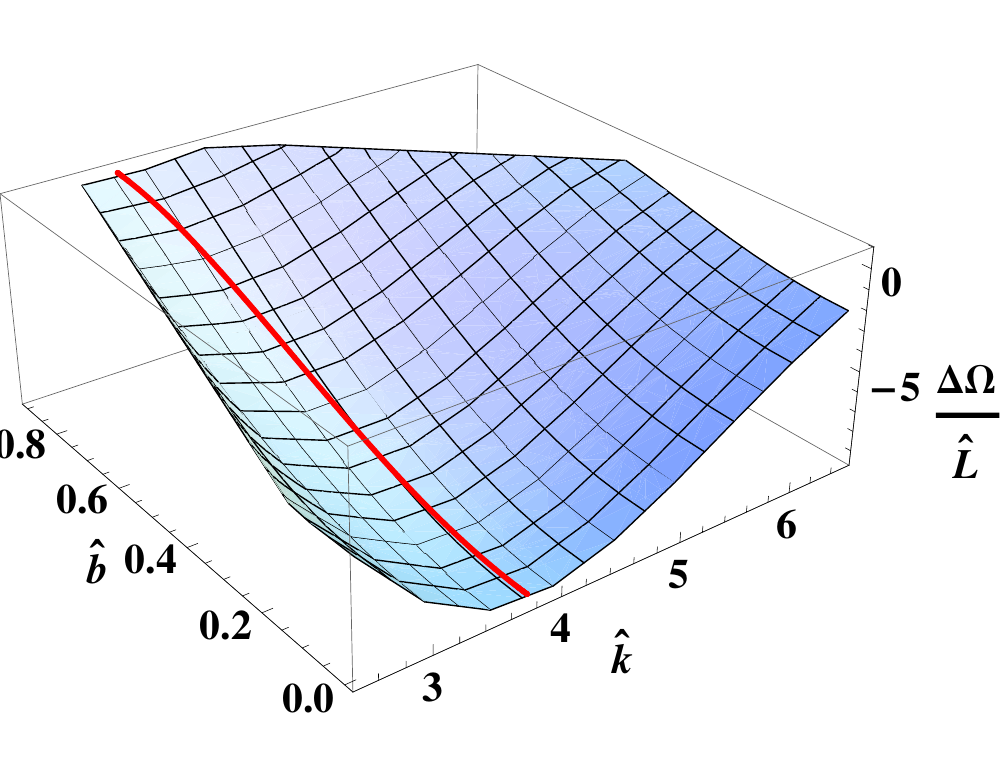}
\caption{$\Delta \Omega/\hat L$ as a function of $\hat b$ and $\hat k$ with $\hat\mu = 4$. The red curve shows $\hat k_0(\hat b)$, which minimizes $\Delta \Omega/\hat L$ for each $\hat b$. The transition from striped to homogeneous occurs at the critical magnetic field $\hat b_c = 0.864$, at the critical frequency $\hat k_c = 3.38$.}
\label{fig:DeltaOmega_vs_k_and_b}
\end{figure}

We now turn the magnetic field back on and again solve the PDEs.  As in the $\hat b = 0$ case, we find solutions with all the fields modulated.  The main qualitative difference is that the $\hat x \to \frac{\hat L}{2} - \hat x$ parity symmetry is now broken.  For each $\hat b$ and with $\hat \mu$ fixed, we again minimize the free energy $\Omega/\hat L$ to find the spatial frequency $\hat k_0$ of the striped phase.  We find that decreasing $\hat b$ causes $\hat k_0$ to increase.  The energy difference $\Delta\Omega/\hat L$ with the homogeneous phase at fixed $\hat\mu$ is plotted as a function of $\hat b$ and $\hat k$ in Fig.~\ref{fig:DeltaOmega_vs_k_and_b}.

\begin{figure}[!ht]
\center
\includegraphics[width=0.7\textwidth]{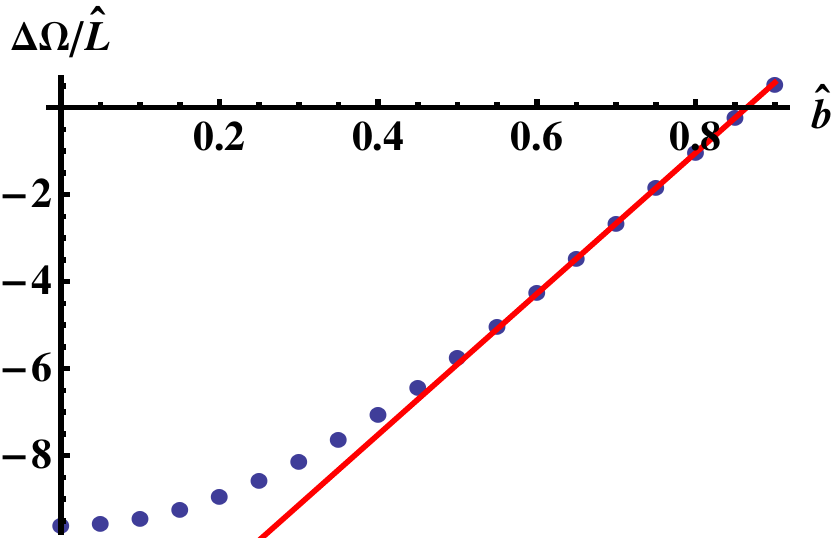}
\caption{The difference in free energy $\Delta \Omega(\hat k_0)/\hat L$ between the homogeneous and striped phases as a function of $\hat b$ with $\hat\mu = 4$.  The blue points are numerically computed values, and the red line is a linear fit $\Delta\Omega/\hat L = 16.2 (\hat b - \hat b_c)$.}
\label{fig:DeltaOmega_with_varying_b}
\end{figure}

We find that the magnetic field suppresses modulation.   For a given $\hat \mu$, modulation is thermodynamically preferred below a critical magnetic field $\hat b_c$, appearing at a critical frequency $\hat k_c > 0$.  Fig.~\ref{fig:DeltaOmega_with_varying_b} shows how the free energy of the striped phase compares with that of the homogeneous phase.  Near $\hat b_c$, the energy difference is a linear function of $\hat b$.  The magnetization $\hat M = -\frac{\partial \Omega}{\partial \hat b}$ therefore jumps at $\hat b_c$,  indicating that the the phase transition is first order, in contrast to the $\hat b = 0$ case; compare  Fig.~\ref{fig:DeltaOmega_with_varying_b} with Fig.~\ref{fig:Omega_and_d_at_zero_b}(left).

\begin{figure}[!ht]
\center
\includegraphics[width=0.7\textwidth]{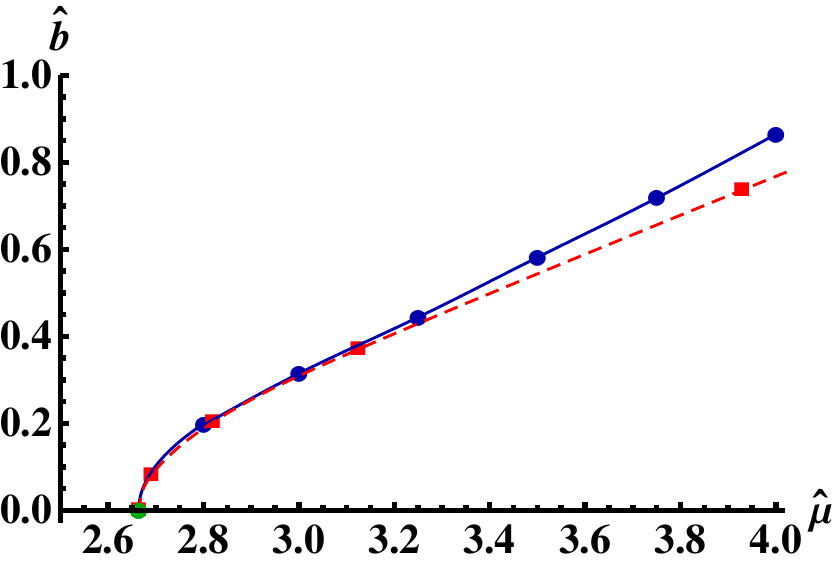}
\caption{The $\hat\mu-\hat b$ phase diagram.  The blue curve $\hat b_c(\hat\mu)$ is a curve of first-order phase transitions between the homogeneous phase (above the curve) and the striped phase (below the curve).  This curve ends in a second-order critical point at $\hat b_c=0$ and $\hat\mu = 2.66$, indicated in green.  The homogeneous phase is perturbatively unstable below the red curve; between the two curves, it is metastable.}
\label{fig:phase_diagram}
\end{figure}

By varying both $\hat \mu$ and $\hat b$, we can map out a two-dimensional phase diagram, which is shown in Fig.~\ref{fig:phase_diagram}.  For each $\hat\mu$, we compute the critical chemical potential $\hat b_c$, which at nonzero $\hat b_c$ is the location of a first-order phase transition.  This line of first-order transitions ends at $\hat b_c = 0$ at a critical point, where the transition becomes continuous.

We can compare this phase diagram Fig.~\ref{fig:phase_diagram} with the results of the fluctuation analysis of the homogeneous phase at nonzero $\hat b$, performed in \cite{Jokela:2012vn}.  Because the phase transition is in general first order, we would expect the homogeneous phase to change only from stable to metastable at the transition point $\hat\mu_c$; the instability should appear at $\hat\mu_{inst} > \hat\mu_c$.  In fact, this is what is found by comparing our results to those of in Sec.~4.2 and Fig.~5 of \cite{Jokela:2012vn}.\footnote{The variables in \cite{Jokela:2012vn} are defined as we do here, but the phase diagram is given in terms of charge density $\hat d$ rather than chemical potential $\hat\mu$.}


\section{Discussion and outlook}
\label{sec:discussion}

In this paper, we presented the striped phase of the D3-D7' system indicated by the perturbative instability of the homogeneous state in \cite{Bergman:2011rf, Jokela:2012vn} and found by numerically solving the coupled equations of motion.  This modulated phase was found to be both a charge density and spin density wave, while the latter component was vastly more dominant.  The phase structure presented in Fig.~\ref{fig:phase_diagram} is in excellent agreement with the previous perturbative analysis.

There have been several recent examples of spontaneously modulated ground states of a similar type as those studied here; in particular, \cite{Withers:2013loa,Withers:2013kva,Donos:2013wia,Rozali:2012es,Rozali:2013ama} analyzed bottom-up holographic models whose homogeneous states suffer from similar types of instability as the D3-D7' system.  They find striped, cohomogeneity-one solutions, dual to charge and/or current density waves. Though these solutions seem to resemble those that we found in this work, our key player was the spin density wave and not so much the charge density wave component of the modulation.  Also, the phase transition from the homogeneous to striped phases in those works were typically found to be of second order, though those systems were not studied in the presence of a magnetic field.

There are a number of interesting avenues left to explore in the future.  The D3-D7' system features a number of parameters which have been fixed so far, such as the mass $m_\psi$ and the internal fluxes $f_1$ and $f_2$, and which could be varied to potentially interesting effect.  In particular, the alternative quantization of the bulk gauge field, studied in \cite{Jokela:2013hta, Jokela:2014wsa}, corresponds to an $SL(2,\mathbb Z)$ mapping of the boundary CFT.  An upcoming stability analysis \cite{upcoming1} indicates that the choice of quantization strongly affects the phase structure, but a construction of modulated solutions will be needed for a complete understanding.

In the quantum Hall phase, corresponding to a Minkowski embedding of the D7-brane, the lowest neutral excitation, for a range of parameters, is a magneto-roton, that is, a collective mode whose minimum energy is at nonzero momentum \cite{Jokela:2010nu}.  One could look for modulated Minkowski solutions, corresponding to a roton condensate.  However, since the homogeneous solution was found to be perturbatively stable, such a modulated state would not likely be preferred energetically.

A compelling but technically difficult open question is whether the striped phase is in fact the ground state. Or, does continuous translation symmetry break completely, resulting in a two-dimensional lattice?  In principle, one could solve the system of PDEs, allowing the fields to depend on both $x$ and $y$.  This presents a challenging numerical problem.\footnote{Recently, in a bottom-up Einstein-Maxwell-scalar model, a solution with two modulated dimensions was found numerically \cite{Withers:2014sja}.} Another, possibly more tractable approach, might be to perform a fluctuation analysis of the striped phase to detect perturbative instabilities.  Just as a tachyon with nonzero momentum signaled an instability of the homogeneous mode to forming stripes, a tachyonic mode with nonzero momentum along the stripes would imply an instability toward forming a lattice.  Such a calculation would involve solving the linearized fluctuation equations on top of the numerical striped background.

Another computation remaining is the calculation of the conductivity of the striped phase.  Translationally invariant phases typically feature infinite DC conductivity, though probe-brane models have finite DC conductivity because momentum can be dissipated into the large $N$ number of bulk degrees of freedom. Still, the conductivity in the modulated direction is expected to have interesting properties.  

There are two standard methods for computing conductivities in holographic probe-brane models.  One can compute the retarded current-current correlator which relates to the conductivity via a Kubo formula.\footnote{For example, see \cite{Brattan:2013wya} where the conductivity is computed in the alternatively quantized D3-D5 model.}  In this case, such a calculation would be similar in difficulty to the linear stability analysis discussed above.  Alternatively, the Karch-O'Bannon method \cite{KOB} can be used to compute the DC conductivity.  However, because the embedding is inhomogeneous and only known numerically, employing this technique will not be straightforward.

Finally, there are many basic open questions to be addressed.  Of particular importance is to what degree symmetry-breaking is generic.  Symmetry-breaking instabilities appear to be a common feature of holographic systems.  To what degree can the result of this model be generalized?  Given the current nature of holographic modeling, the detailed features of any one construction is less relevant than generic features shared by a wide class of systems.

For example, the magnetic field tends to inhibit the symmetry-breaking instability in the D3-D7' system.  A similar effect was also seen in the linear stability analysis of the closely related D2-D8' model \cite{Jokela:2012se}, as well as in the Sakai-Sugimoto model \cite{BallonBayona:2012wx}.  We might be tempted to conjecture that magnetic fields tend to stabilize modulated instabilities more generally.  This type of statement is about as much as we can currently aim for in holographic condensed matter physics.
The instabilities in these models are all of the type identified in \cite{Nakamura:2009tf}, so this may fail generalize to instabilities generated by different mechanisms.  So, it would be interesting to investigate the stabilizing effects of a magnetic field in other holographic models, to test this conjecture.

\vspace{0.5cm}

{\bf \large Acknowledgments}
We thank Francesco Aprile, Daniel Arean, Takaaki Ishii, Gilad Lifschytz, Javier Mas, Christiana Pantelidou, and Ben Withers for discussions.
N.J. is supported by the Academy of Finland grant no. 1268023. M.J. is partially supported by European Union's Seventh Framework Programme
under grant agreements (FP7-REGPOT-2012-2013-1) No 316165, PIF-GA-
2011-300984, the ERC Advanced Grant BSMOXFORD 228169, the EU program
Thales MIS 375734. He is also co-financed by the European Union (European Social
Fund, ESF) and Greek national funds through the Operational Program ``Education
and Lifelong Learning'' of the National Strategic Reference Framework (NSRF)
under ``Funding of proposals that have received a positive evaluation in the 3rd and
4th Call of ERC Grant Schemes'', as well as under the action ``ARISTEIA''. 
M.L. is supported by funding from the European Research
Council under the European Union's Seventh Framework Programme (FP7/2007-2013) /
ERC Grant agreement no.~268088-EMERGRAV.  
This work is part of the $\Delta$-ITP consortium, a program of the Netherlands Organisation for Scientific Research (NWO) that is funded by the Dutch Ministry of Education, Culture and Science (OCW).
In addition, we acknowledge extensive use of the computational resources of CCTP.  And, finally, we thank the ESF Holograv Network and $\Delta$-ITP for supporting the ``Workshop on Holographic Inhomogeneities", during which this work was finalized.

\appendix

\section{Finding inhomogeneous solutions numerically}
\label{app:method}

In this appendix, we describe in more detail the pseudospectral method for numerically solving the equations of motion. 

\subsection{Boundary conditions}
\label{app:boundaryconditions}

Let us start from the boundary conditions in the UV, where $u \to 0$. After using the radial gauge condition and rotation symmetry in the $x$-$y$-plane, the dynamical fields are $\psi$, $\hat z$, $\hat a_{0}$, and $\hat a_{y}$.  One can solve the equations of motion as a series around $u=0$, assuming unmodulated sources:
\bea
\label{psiUVexpansion}
\psi(\hat x,u) &=& \psi_\infty + \hat m_\psi u - \hat c_\psi(\hat x) u^2+\frac{55}{6}\hat m_\psi^3u^3 + \nonumber\\
&&+\frac{1}{48} \left(-480 \hat m_\psi^2 \hat c_\psi(\hat x)-16\hat b\hat d(x)+9  \hat c_\psi''(\hat x)\right)
u^4 + \cdots \\
\label{zUVexpansion}
\hat z(\hat x,u) &=& \bar z+\frac{1}{2\sqrt{2}}  u+\frac{\hat m_\psi^2}{4\sqrt{2}}u^3-\frac{\hat m_\psi}{2\sqrt{2}} \hat c_\psi(\hat x)u^4+ \nonumber\\
&&+\hat c_z(\hat x)u^5-\frac{320 \hat m_\psi^3 \hat c_\psi(\hat x)-3 \hat m_\psi \hat c_\psi''(\hat x)}{32 \sqrt{2}} u^6 + \cdots \\
\label{a0UVexpansion}
\hat a_{0}(\hat x,u) &=& \hat \mu - \frac{\hat d(\hat x)}{\sqrt{2}} u+\sqrt{2}\hat b \hat m_\psi u^2 - \nonumber\\
&&-\frac{64 \hat b \hat c_\psi(\hat x)+24 \hat m_\psi^2 \hat d(\hat x)-\hat m_\psi \hat j_y'(\hat x)-9 \hat d''(\hat x)}{48 \sqrt{2}}
u^3 +\cdots \\
\label{ayUVexpansion}
\hat a_{y}(\hat x,u) &=& \bar a_y + \hat b\hat x + \hat j_y(\hat x)u+
\frac{1}{48} \left(24 \hat m_\psi^2 \hat j_y(\hat x)-16 \hat m_\psi \hat d'(\hat x)-9 \hat j_y''(\hat x)\right)
u^3 + \cdots
\eea

The constants $\hat m_\psi$, $\hat \mu$, and $\hat b$ are parameters we are free to choose.  
The translation symmetry of the background transverse to the D7-brane means that the constants $\bar z$ and $\bar a_y$ can be set to any value without affecting observables, so we can set them to zero without loss of generality.   
Therefore, we impose the following UV boundary conditions:
\bea \label{eq:UVbcfirst}
 \psi(\hat x,0) &=& \psi_\infty \\
 \partial_u \psi(\hat x,0) &=& \hat m_\psi \\
 \hat a_{0}(\hat x,0) &=& \hat\mu\\
 \hat z(\hat x,0) &=& 0 \\
  \hat a_{y}(\hat x,0) &=& \hat b\hat x \ . \label{eq:UVbclast}
\eea
These free parameters have clear physical interpretations. The $\hat\mu$ is the (reduced) chemical potential for the fermions of mass $\hat m_\psi$, and $\hat b$ is the (reduced) external magnetic field:
\bea
 \hat m_\psi & = & \frac{m_\psi}{r_T} \\ 
 \hat\mu & = & \frac{\mu}{r_T} \\
 \hat b & = & \frac{b}{r^2_T} \ .
\eea

In the IR, at the horizon where $u \to 1$, we require that the solutions are regular and finite. In particular, we notice that the action \eqref{eq:uaction} contains, in $\hat A_{x}$ of \eqref{Axexp}, a term $\frac{1}{h}\partial_{\hat x}  \hat a_{0}^2$.  Since $h \to 0$ at the horizon, we need to require that $\partial_{\hat x}  \hat a_{0} \to 0$ in order for the action to remain real. Therefore, $\hat a_{0}$ takes a constant value on the horizon, which must be set to zero in order for $\hat a$ to be a well-defined one-form at the horizon.  This choice also implies that the boundary value of $\hat a_0$ is precisely the (reduced) chemical potential. The value of $\hat a_{0}$ is the only integration constant we need to fix in the IR, as the behavior of the various fields is otherwise determined by requiring regularity. It turns out, however, that imposing explicitly the leading-order constraint, which follow from expanding the equations of motion as series at $u=1$, improves the stability of the code.  We 
therefore require explicitly that 
our solution satisfies both
\be \label{eq:a0IR}
 \hat a_{0}(\hat x,1) = 0 
\ee
as well as the three leading-order conditions from the equations of motion for the fields $\psi$, $\hat z$, and $\hat a_{y}$, which are lengthy expressions and not reproduced here.

Note that the action \eqref{eq:uaction} is covariant under the reflection $\hat x \to -\hat x$ if $\hat a_{y}$ is odd and the other fields even in $\hat x$.\footnote{The discrete symmetry properties of the D3-D7' model were studied in detail in  \cite{Omid:2012vy}.} This property is reflected in the UV expansions \eqref{psiUVexpansion}, \eqref{zUVexpansion}, \eqref{a0UVexpansion}, and \eqref{ayUVexpansion} (if we set to zero the ``trivial'' constant $\bar a_y$). When $\hat b=0$, there is also another symmetry: the action is covariant under the reflection if $\psi-\pi/4$ is odd and the other fields are even in $\hat x$. There does not appear to be a specific reason solutions must obey this parity behavior.  But, any inhomogeneous solutions we found respected the first symmetry and the solutions with $\hat b=0$ respected both symmetries when reflected with respect to some value of $\hat x$. By translation invariance we required that the first symmetry appears in reflections with respect to  $\hat x=0$. Then 
the 
solutions with
$\hat b=0$ have the second symmetry in reflections with respect to $\hat x = \hat L/4$.

\subsection{Implementation of the pseudospectral method}
\label{app:expansion}

Having fixed the boundary conditions,  
we implement the pseudospectral method based on a Fourier series in the $x$-direction and an expansion in the Chebyshev polynomials in the $u$-direction. 
We assume that the desired inhomogeneous solution respects the first parity symmetry with respect to $\hat x=0$.\footnote{We tried relaxing this assumption but did not find additional solutions.} Therefore, we choose the collocation points from a grid of $N_x+1$ evenly spaced points ranging from $\hat x=0$ to $\hat x=\hat L/2$ in the $x$-direction and from the Gauss-Lobatto grid with $N_u$ points ranging from $u=0$ to $u=1$ in the $u$-direction. The problem is then reduced to finding the values of all the functions $\psi$, $\hat z$, $\hat a_{0}$, and $\hat a_{y}$ at the collocation points. 

The solution is found by requiring that the boundary conditions and equations of motion are satisfied at the collocation points. Notice that the UV boundary conditions \eqref{eq:UVbcfirst}--\eqref{eq:UVbclast}, except for the second one fixing the quark mass, and the IR boundary condition~\eqref{eq:a0IR} directly fix the values of the functions at the boundary points, thus reducing the number of variables. For the other conditions, we need the values of the derivatives at collocation points. These are computed from the values of the functions in the pseudospectral approximation. In more detail, we require that the values of the functions match with the expansions
\bea
 \psi(\hat x,u) &=& \sum_{j=0}^{N_u-1}\sum_{k=0}^{N_x-1}  \hat \psi_{j,k} T_{j}(2u-1) \cos\frac{2\pi k \hat x}{\hat L} \\
 \hat a_{0}(\hat x,u) &=& \sum_{j=0}^{N_u-1}\sum_{k=0}^{N_x-1}  \hat a_{0,j,k} T_{j}(2u-1) \cos\frac{2\pi k \hat x}{\hat L} \\
 \hat z(\hat x,u) &=& \sum_{j=0}^{N_u-1}\sum_{k=0}^{N_x-1}  \hat z_{j,k} T_{j}(2u-1) \cos\frac{2\pi k \hat x}{\hat L} \\
 \hat a_{y}(\hat x,u) &=& \sum_{j=0}^{N_u-1}\sum_{k=1}^{N_x-1}  \hat a_{y,j,k} T_{j}(2u-1) \sin\frac{2\pi k \hat x}{\hat L} 
\eea
at the collocation points. Here $T_j$ are Chebyshev polynomials, defined in the range $[-1,1]$. This fixes the coefficients  $\psi_{j,k}$, $\hat z_{j,k}$, $\hat a_{0,j,k}$, and $\hat a_{y,j,k}$ in terms of the values of the functions. The derivatives are then computed from these expansions. 

Using the expressions for the derivatives, the remaining boundary conditions and equations of motions can be turned into algebraic conditions for the values of the functions. The UV and IR boundary conditions are evaluated at the collocation points at $u=0$ and at $u=1$, respectively, and the four equations of motion are evaluated at the other collocation points with $0<u<1$, so that the total number of conditions matches with the number of remaining variables. The resulting algebraic equations are then solved numerically by using the Newton method.

\subsection{Initial configurations for the Newton method}

The last step in the numerical method, {\it i.e.}, solving the algebraic conditions discussed above, is a rather nontrivial task for our system. 
The difficulties arise because unlike in most AdS/CMT models where inhomogeneities have been studied in the literature, the 
square roots in our brane action give rise to nonanalytic behavior. As it turns out, the inhomogeneous solutions typically lie close to the branch points of the square roots and 
tiny deformations of these solutions can make the action complex. Consequently, the initial guess for the Newton method must be chosen very carefully, as otherwise the square root factors become complex at some intermediate step, which causes the method to diverge.

Naturally, the main idea for constructing initial conditions is to start at low $N_x$ and $N_u$, and iteratively increase the number of grid points using the previous solution as initial guess for the next step. In our case, however, the inhomogeneous solution may be absent at very low number of grid points (or more precisely, the solution has turned complex), which limits the range of $N_x$ and $N_u$ where the iterative procedure can be applied.

In order to overcome these difficulties, we used a separate code for generating the initial data, which also relied on a spectral method. We started with a very small grid (3-4 points in each direction) where the system could be solved by an initial Ansatz motivated by the fluctuation analysis~\cite{Bergman:2011rf,Jokela:2012vn}. The existence of the solution was guaranteed by adding a small random noise in the locations of the collocation points and by running the code in loop until it found a valid real root. The size of the grid was then gradually increased, keeping the random noise, until the grid was about ten points in each direction. After this, the obtained solution could be used as an initial configuration for the actual code, and grid size could be iteratively increased further without issues.

\subsection{Convergence and accuracy}

We tested the convergence of the code at a single set of parameter values ($\hat \mu=4$, $\hat L =\pi/2$, $\hat m=0$, and $\hat b=0$) up to $N_x=38$ and $N_u=40$. The expected roughly exponential convergence was found. At the largest grid size, the relative accuracy of $\hat a_{y}$ was about $10^{-9}$, that of $\psi$ and $\hat a_{0}$ was about $10^{-10}$, and that of $\hat z$ was about $10^{-11}$. Most of the data in this article was computed using $N_x=24$ and $N_u=26$. At this grid size, the accuracy of  $\hat a_{y}$ was smaller than $10^{-6}$, that of $\psi$ and $\hat a_{0}$ was smaller than $10^{-7}$, and that of $\hat z$ was smaller than $10^{-8}$. This accuracy is more than enough for all analysis presented here.


\end{document}